# General Electrical Manipulations of Electron Spin

Z. Y. Wang[*], C. D. Xiong

*University of Electronic Science and Technology of China, Chengdu, Sichuan 610054, P.R. China*

**Abstract**

Traditionally, the interactions related to the 3D spatial angular momentum have been studied completely, while the ones related to the generators of Lorentz boost are always ignored. In this paper we show that the generators of Lorentz boost have a nontrivial physical significance in quantum mechanics, and try to propose a most general theory about electrical manipulations of electron spin, where a new treatment and interpretation for the traditional Darwin term and spin-orbit coupling is given via the concept of electron's *induced* electric moment. Some *electrostatic* fields can even affect the spin quantum states of a *resting* electron, and parallel electric and magnetic fields can be simultaneously applied to fix on the spin orientation of electron.

[*]Electronic address: zywang@uestc.edu.cn

1. **Introduction**

On the one hand, for a closed system, the conservation of a Lorentz-boost vector only implies that the velocity of the relativistic inertia-center of the system is a constant [1-2]; on the other hand, in general, the infinitesimal generators of Lorentz boost are not conserved, which is why we do not use its eigenvalues to label physical states [3]. Therefore, the physical significances of the generators of Lorentz boosts seem to be *trivial*. As a result of which, traditionally, people only consider and study the generators of 3D spatial rotation, i.e., the 3D spatial angular momentum, while the generators of Lorentz boost are always ignored.

However, in this paper we shall show that the generators of Lorentz boost have a nontrivial physical significance in quantum mechanics, and there do have interactions related to the generators of Lorentz boost. In fact, in **Appendix A** we show that, to a charged system with nonzero space-time components of 4D angular momentum tensor (i.e. the generators of Lorentz boost) there may correspond to an electric moment, just as that, to a charged system with nonzero purely spatial components of 4D angular momentum tensor (i.e. the generators of 3D spatial rotation) there may correspond to a magnetic moment. The potential technology significance of the interactions related to the generators of Lorentz boost lies in that, it may open attractive and multiplicate applications in future electronic devices with smaller size. For example, this may offer other ways to strengthen the control and manipulation for the spin degrees of freedom of electron for *spintronics*.

For simplicity we apply natural units of measurement ($\hbar = c = 1$) and adopt the same notation as Ref. [4] except additional indication.

## 2. 4D spin-orbit tensor coupling of electron

Consider an infinitesimal proper Lorentz transformation that changes the 4D space-time coordinates according to $x^\mu \to x'^\mu = (g^{\mu\nu} + \varepsilon^{\mu\nu})x_\nu$ ( $\mu,\nu = 0,1,2,3$ ), where $\varepsilon^{\mu\nu}$ as an infinitesimal antisymmetric tensor denotes the infinitesimal boost parameters or rotation angles, $g_{\mu\nu} = diag(1,-1,-1,-1)$ is the metric tensor. Owing to the Lorentz symmetry, a $n$-component field $\psi(x) = (\psi^1,...,\psi^n)^T$ (the superscript T denotes matrix transpose) is transformed as

$$\psi(x) \to \psi'(x') = (1 - \frac{i}{2}\varepsilon_{\mu\nu}S^{\mu\nu})\psi(x) \qquad (1)$$

where $S^{\mu\nu} = -S^{\nu\mu}$ are the infinitesimal generators of the Lorentz transformation, also the 4D spin tensor of $\psi(x)$. The corresponding total variation of $\psi(x)$ is

$$\delta\psi = \psi'(x) - \psi(x) = -\frac{i}{2}\varepsilon_{\mu\nu}\hat{J}^{\mu\nu}\psi(x) \qquad (2)$$

where

$$\hat{J}^{\mu\nu} = x^\mu i\partial^\nu - x^\nu i\partial^\mu + S^{\mu\nu} = \hat{L}^{\mu\nu} + S^{\mu\nu} \qquad (3)$$

is the tensor of 4D total angular momentum with $\hat{L}^{\mu\nu} = x^\mu \hat{p}^\nu - x^\nu \hat{p}^\mu$ the tensor of 4D orbital angular momentum, here $\hat{p}_\mu = i\partial/\partial x^\mu \equiv i\partial_\mu$ denotes the 4-momentum operator (while $\hat{p} = -i\nabla$ the 3-momentum operator). The operators $\hat{J}^{\mu\nu}$ are also called the generators of the Lorentz transformation with $S^{\mu\nu}$ the *infinitesimal* generators. As we known, the space components of $\hat{J}^{\mu\nu}$ or $S^{\mu\nu}$ correspond to the generators of spatial rotations, and can be mapped to the 3D angular-momentum vector; while the components with mixed spatial-temporal indices correspond to the generators of Lorentz boosts, and can also be mapped to a 3D vector, we call it the Lorentz-boost vector.

Now, let the field (or wave function) $\psi(x)$ be the Dirac field of electron, then all the discussions above are valid for the Dirac field. Let $A^\mu = (\Phi, \vec{A})$ ($\mu = 0,1,2,3$) be the 4-potential of an external electromagnetic field ($\vec{A}$ being the vector potential and $A^0 = \Phi$ the scalar potential), $e$ the unit charge, and $F_{\mu\nu} \equiv \partial_\mu A_\nu - \partial_\nu A_\mu$ the electromagnetic field tensor. In general, we define the electric and magnetic field strengths $\vec{E} \equiv -\nabla\Phi - \frac{\partial}{\partial t}\vec{A}$, $\vec{B} \equiv \nabla \times \vec{A}$, they are assembled into a covariant anti-symmetric tensor, i.e. $F_{\mu\nu}$. Let $D_\mu \equiv \partial_\mu + ieA_\mu$ denote the gauge covariant derivative, the Dirac equation of electron in $A^\mu$ is

$$(i\gamma^\mu D_\mu - m)\psi(x) = 0 \qquad (4)$$

where the four $4\times 4$ Dirac matrices $\gamma^\mu$ satisfy the algebra $\gamma^\mu\gamma^\nu + \gamma^\nu\gamma^\mu = 2g^{\mu\nu}$, the 4-component wavefunction $\psi(x)$ satisfies the transformation laws of a relativistic spinor. In our case, the infinitesimal generators of the Lorentz transformation are $S^{\mu\nu} = \frac{i}{4}[\gamma^\mu, \gamma^\nu]$, i.e., the 4D spin tensor of electron. Let $\varepsilon^{ijk}$ denotes the totally anti-symmetric tensor with $\varepsilon^{123} = 1$ ($i,j,k = 1,2,3$), one can show that $\vec{\Sigma} = (\Sigma_1, \Sigma_2, \Sigma_3)$ with $\Sigma_i = \frac{1}{2}\varepsilon_{ijk}S^{jk}$ is the usual spin matrices (as the generators of 3D spatial rotations). We call the generators $\vec{K} \equiv (S^{01}, S^{02}, S^{03})$ of Lorentz boosts as *spin-like matrices*. Let $\vec{\sigma} = (\sigma_1, \sigma_2, \sigma_3)$ be the Pauli's matrix vector, where

$$\sigma_1 = \begin{pmatrix} 0 & 1 \\ 1 & 0 \end{pmatrix}, \quad \sigma_2 = \begin{pmatrix} 0 & -i \\ i & 0 \end{pmatrix}, \quad \sigma_3 = \begin{pmatrix} 1 & 0 \\ 0 & -1 \end{pmatrix} \qquad (5)$$

In terms of $\vec{\sigma}$, the spin matrices $\vec{\Sigma}$ and *spin-like* matrices $\vec{K}$ can be expressed as

$$\vec{\Sigma} = \frac{1}{2}\begin{pmatrix} \vec{\sigma} & 0 \\ 0 & \vec{\sigma} \end{pmatrix}, \quad \vec{K} = \frac{i}{2}\begin{pmatrix} 0 & \vec{\sigma} \\ \vec{\sigma} & 0 \end{pmatrix} \qquad (6)$$

Note that $\vec{\alpha} = \begin{pmatrix} 0 & \vec{\sigma} \\ \vec{\sigma} & 0 \end{pmatrix} = -2i\vec{K}$ also plays the role of the velocity operator of electron [5]. The first-order Dirac equation (4) can be transformed into second-order one [6]:

$$[(i\partial^\mu - eA^\mu)(i\partial_\mu - eA_\mu) - m^2 - eS^{\mu\nu}F_{\mu\nu}]\psi(x) = 0 \qquad (7)$$

With respect to Equ. (7), in contrast with the traditional treatment, using $\psi = \frac{i\gamma^\mu D_\mu}{m}\psi$ obtained from Equ. (4), we write the last term on the light-hand side of Equ. (7) as

$$-eS^{\alpha\beta}F_{\alpha\beta}\psi = -\frac{e}{m}S^{\alpha\beta}F_{\alpha\beta}[i\gamma^\lambda D_\lambda]\psi \qquad (\alpha, \beta, \lambda = 0,1,2,3) \qquad (8)$$

Consider that $\partial_\mu = \frac{\partial}{\partial x^\mu} = \frac{x_\mu}{\tau}\frac{\partial}{\partial \tau}$, where $\tau = \sqrt{x_0^2 - \vec{x}^2} = \sqrt{x^\mu x_\mu}$ is the proper time, one has $F_{\mu\nu} = \partial_\mu A_\nu - \partial_\nu A_\mu = \frac{x_\mu}{\tau}\frac{\partial A_\nu}{\partial \tau} - \frac{x_\nu}{\tau}\frac{\partial A_\mu}{\partial \tau}$ (in the presence of the interactions between electromagnetic waves and electric charges, e.g., the electromagnetic waves propagate in media, $\tau \neq 0$ in general). Using $a^\mu\gamma_\mu b^\nu\gamma_\nu = a^\mu b_\mu - 2ia^\mu b^\nu S_{\mu\nu}$, $S^{\mu\nu} = i[\gamma^\mu, \gamma^\nu]/4$, $D_\mu = \partial_\mu + ieA_\mu$ and the Lorentz gauge condition $\partial_\mu A^\mu = 0$, we obtain

$$-eS^{\alpha\beta}F_{\alpha\beta}\psi = [-\frac{e}{m}\gamma_\mu(\partial_\nu A^\mu)D^\nu + \frac{e}{m}\kappa S_{\alpha\beta}\hat{L}_A^{\alpha\beta}]\psi \qquad (9)$$

where $\hat{L}_A^{\mu\nu} \equiv x^\mu iD^\nu - x^\nu iD^\mu$ is the 4D orbit-angular-momentum tensor in the external fields $A^\mu$ $\kappa \equiv \frac{1}{\tau}\frac{\gamma^\mu \partial A_\mu}{\partial \tau}$ is a matrix parameter. Then Equ. (7) becomes

$$[D^\mu D_\mu - m^2 - \frac{e}{m}\gamma_\mu(\partial_\nu A^\mu)D^\nu + \frac{e}{m}\kappa S_{\alpha\beta}\hat{L}_A^{\alpha\beta}]\psi(x) = 0 \qquad (10)$$

Now let $\psi = \psi'\exp(-im)$, in the approximation $|i\partial\psi'/\partial t|$, $|e\Phi\psi'| \ll m|\psi'|$ and $|i\partial\Phi/\partial t| \ll m|\Phi|$, Equ. (10) becomes

$$i\frac{\partial}{\partial t}\psi' = [\frac{(\hat{p}-e\vec{A})^2}{2m} + e\Phi + \frac{e}{2m^2}\gamma_\mu(\partial_\nu A^\mu)D^\nu - \frac{e}{2m^2}\kappa S_{\alpha\beta}\hat{L}_A^{\alpha\beta}]\psi' \quad (11)$$

Equ. (10)-(11) are for the first time obtained by us. We refer to $\frac{e}{2m^2}\kappa S_{\alpha\beta}\hat{L}_A^{\alpha\beta}$ in Equ. (11) as *the 4D spin-orbit tensor coupling*, which describes the coupling between the 4D spin and orbit-angular-momentum tensors, and includes not only the usual 3D spatial spin-orbit coupling but also the interaction related to the Lorentz boost vectors. For example, let $\Phi = \Phi(r)$, $r = \sqrt{\vec{x}^2 + \vec{y}^2 + \vec{z}^2}$, $\hat{L} = \vec{r}\times\hat{p}$ denotes the 3D spatial angular momentum and the electric field strength $\vec{E} = -\nabla\Phi(r) = -\frac{\vec{r}}{r}\frac{d}{dr}\Phi(r)$, consider that $\psi = \frac{i\gamma^\mu D_\mu}{m}\psi$, the last two terms on the right-hand side of Equ. (11) become, approximatively,

$$[-\frac{e}{m}\gamma^0(\vec{\Sigma}\cdot\vec{B}) - \frac{ie}{4m^2}\vec{\Sigma}\cdot(\nabla\times\vec{E}) - \frac{e}{8m^2}\nabla\cdot\vec{E} - \frac{e}{2m^2}\frac{1}{r}\frac{\partial\Phi}{\partial r}(\vec{\Sigma}\cdot\hat{L})]\psi' \quad (12)$$

where the third term on the right-hand side of Equ. (12) is the so-called Darwin term and last term represents the usual spin-orbit coupling. By Equ. (10), starting from a different approximation, one will obtain a different result.

Therefore, to the Dirac equations of electron in external fields, there contain the coupling between the 4D spin tensor and the 4D orbit-angular-momentum tensor, with the interactions related to the Lorentz boost vectors being included.

### 3. *spin-like* degrees of freedom of electron

By virtue of the concepts of *spin-like* degrees of freedom and *induced* electric moment of electron, we will reveal other physical meanings for the infinitesimal generators of Lorentz boost and give a physical interpretation for them.

In term of the electric and magnetic field strengths $\vec{E} \equiv -\nabla\Phi - \frac{\partial}{\partial t}\vec{A}$, $\vec{B} \equiv \nabla\times\vec{A}$,

Equ. (11) can be rewritten as

$$i\frac{\partial}{\partial t}\psi' = [\frac{(\hat{p}-e\vec{A})^2}{2m} + e\Phi - g_s\frac{e}{2m}\vec{\Sigma}\cdot\vec{B} + g_s\frac{e}{2m}\vec{K}\cdot\vec{E}]\psi' \quad (13)$$

where the g-factor $g_s = 2$. The $i=\sqrt{-1}$ factor in $\vec{K}$ (expressed by Equ. (6)) is necessary to assure the Hermiticity of $\vec{K}\cdot\vec{E}$ in Equ. (13). The electron will therefore behave as though it has a magnetic moment operator $g_s\frac{e}{2m}\vec{\Sigma}$ and an electric moment operator $g_s\frac{e}{2m}\vec{K}$. As well known, an observable is represented by the expectation value of a dynamical operator or the squared norm of a state vector, instead of by the dynamical operator or state vector itself. Especially, in the Schrödinger picture, the relativistic effects of the purely numerical matrix operators (such as $\vec{\Sigma}$ and $\vec{K}$) are carried via the wave functions (or state vectors) that operated by these operators. Judged by the relation between the averages of $g_s\frac{e}{2m}\vec{\Sigma}$ and $g_s\frac{e}{2m}\vec{K}$ in the same wave function $\psi'$, one can show that the magnetic moment, related to the spin matrices $\vec{\Sigma}$, is nonzero in the rest frame of electron (say, the magnetic moment is *intrinsic*); whereas the electric moment, related to *spin-like* matrices $\vec{K}$, is the relativistic effect of the magnetic moment (say, the electric moment is *induced*). For example, let $\psi' = \begin{pmatrix}\varphi'\\\chi'\end{pmatrix}$, i.e., the four-component spinor $\psi'$ is decomposed into two two-component spinors $\varphi'$ and $\chi'$, using Equ. (13) one can show that, in the nonrelativistic and weak external fields limit, $(\vec{\Sigma}\cdot\vec{B}-\vec{K}\cdot\vec{E})\psi' \to (\vec{\sigma}\cdot\vec{B}-\frac{v}{c}\vec{\sigma}\cdot\vec{E})\varphi'$, where $v$ and $c(=1)$ are the velocities of the electron and light in the vacuum, respectively.

In a word, just as that the spin degrees of freedom related to the infinitesimal generators of spatial rotation (i.e., the spin matrices $\bar{\Sigma}$) bring an *intrinsic* magnetic moment for electron, the *spin-like degrees of freedom* related to the infinitesimal generators of Lorentz boost (i.e., the *spin-like* matrices $\bar{K}$) bring an *induced* electric moment for electron.

Then, how to understand the physical meanings of electron's *induced* electric moment and the *like-spin* degrees of freedom? As we known, for the wavepacket $\psi = \begin{pmatrix} \varphi \\ \chi \end{pmatrix}$ of *negative* electron, $\varphi$ is the large component and $\chi$ the small component, while for that of *positive* electron, $\chi$ is the large component and $\varphi$ the small component. In **Appendix B** we show that, when the wavepacket $\begin{pmatrix} \varphi \\ \chi \end{pmatrix}$ of electron is moving with high speeds or varies rapidly, or equivalently its size is sufficiently small, or in the present of a strong external field, the small components $\chi$ of $\begin{pmatrix} \varphi \\ \chi \end{pmatrix}$ can not be ignored. For the moment the wavepacket is constituted by a superposition of positive- and negative-energy components, and the relative intensities of them are proportional to $|\varphi|^2 / |\chi|^2$. Consider that a negative-energy solution corresponds to a positive-electron one, thus this implies that the negative-electron wavepacket does also contain a positive-electron *component* (the latter as the relativistic effect of the former is a small component), and vice versa, just as mentioned in **Appendix B**. When the wavepacket of an electron contains a small positive-electron component, its *induced* electric moment is nonvanishing, and this belongs to a pure relativistic-quantum-mechanics effect that has not a counterpart in classical mechanics.

Likewise, due to the fact that a superposition of plane waves of positive- as well as of negative-energy is necessary to obtain a wave packet of electron [7], in contrast to the spin degrees of freedom that related to the infinitesimal generators of spatial rotations (i.e., the spin matrices $\vec{\Sigma}$) and described by spin-↑ and spin-↓ states, the *spin-like* degrees of freedom that related to the infinitesimal generators of Lorentz boosts (i.e., the *spin-like* matrices $\vec{K}$), belong to *the particle-antiparticle space* spanned by the positive- and negative-energy states of electron (the two *spin-like* states can form another different basis for *the particle-antiparticle space*). In other words, the simultaneously presence of the spin and spin-like degrees of freedom implies the wave function $\psi(x)$ of electron has four components rather than two components.

Let us emphasize that the so-called *intrinsic* or *permanent* dipole electric moment of electron [8] that people are searching nowadays (in order to provide an interpretation for the violation of time reversal invariance) is conceptually distinct from the *induced* electric moment of electron mentioned here. The *intrinsic* electric dipole moment implies a charge distribution with respect to the center of a particle, while the *induced* electric moment, related to the infinitesimal generators of Lorentz boost, originates from the relativistic-quantum-mechanics effects that have not a counterpart in classical mechanics. The interaction between the *induced* electric moment and electric fields does not break time reversal symmetry.

**4. Interactions related to the generators of Lorentz boost**

In section 2, we have show that there do have interactions related to the generators of Lorentz boost, which is contained in the coupling between the 4D spin and the orbit-angular-momentum tensors. Now we give another more specific illustration for the interactions related to the generators of Lorentz boost. By applying $\psi' = \begin{pmatrix} \varphi' \\ \chi' \end{pmatrix}$, Equ. (13) becomes

$$i\frac{\partial}{\partial t}\varphi' = [\frac{(\hat{p}-e\vec{A})^2}{2m} + e\Phi - \frac{e}{2m}\vec{\sigma}\cdot\vec{B}]\varphi' + i\frac{e}{2m}\vec{\sigma}\cdot\vec{E}\chi'$$
$$i\frac{\partial}{\partial t}\chi' = [\frac{(\hat{p}-e\vec{A})^2}{2m} + e\Phi - \frac{e}{2m}\vec{\sigma}\cdot\vec{B}]\chi' + i\frac{e}{2m}\vec{\sigma}\cdot\vec{E}\varphi'$$
(14)

One can easily show the Hermiticity of $i\frac{e}{2m}\vec{\sigma}\cdot\vec{E}$ in Equ. (14) because both the $\vec{E}$ and $i\vec{\sigma}$ operators are anti-Hermitian and $[\vec{E}, i\vec{\sigma}] = 0$. Then the interaction $i\frac{e}{2m}\vec{\sigma}\cdot\vec{E}$, related to *spin-like* degrees of freedom, results in the coupling of two-component spinors $\varphi'$ and $\chi'$. One can easily show that the spin-orbit coupling and the Darwin interactions are contained in $i\frac{e}{2m}\vec{\sigma}\cdot\vec{E}$.

Let $\psi'_+ = \varphi' + \chi'$ and $\psi'_- = \varphi' - \chi'$, Equ. (14) becomes

$$i\frac{\partial}{\partial t}\begin{pmatrix}\psi'_+ \\ \psi'_-\end{pmatrix} = \hat{H}\begin{pmatrix}\psi'_+ \\ \psi'_-\end{pmatrix}$$
(15)

Where

$$\hat{H} = \begin{pmatrix} \hat{H}_0 + i\frac{e}{2m}\vec{\sigma}\cdot\vec{E} & 0 \\ 0 & \hat{H}_0 - i\frac{e}{2m}\vec{\sigma}\cdot\vec{E} \end{pmatrix}, \quad \hat{H}_0 = \frac{(\hat{p}-e\vec{A})^2}{2m} + e\Phi - \frac{e}{2m}\vec{\sigma}\cdot\vec{B} \quad (16)$$

For simplicity, we consider a special case: let $\vec{B} = \nabla\times\vec{A} = 0$, $\vec{E} = -\nabla\Phi - \frac{\partial}{\partial t}\vec{A}$ with $\frac{\partial\vec{E}}{\partial t} = 0$, and $\hat{H}_0 = \frac{(e\vec{A})^2}{2m} + e\Phi$ (namely, a *resting* electron in a purely *electrostatic* field), then $[\hat{H}_0, i\frac{e}{2m}\vec{\sigma}\cdot\vec{E}] = 0$, this means that $\hat{H}_0$ and $i\frac{e}{2m}\vec{\sigma}\cdot\vec{E}$ can be diagonalized together and have common eigenstates, let

$$\hat{H}_0 \begin{pmatrix} \psi'_+ \\ \psi'_- \end{pmatrix} = \varepsilon_0 \begin{pmatrix} \psi'_+ \\ \psi'_- \end{pmatrix} \quad i\frac{e}{2m}\vec{\sigma}\cdot\vec{E} \begin{pmatrix} \psi'_+ \\ \psi'_- \end{pmatrix} = \varepsilon \begin{pmatrix} \psi'_+ \\ \psi'_- \end{pmatrix} \quad (17)$$

where we let $\varepsilon > 0$ without loss of generality, then

$$\hat{H} \begin{pmatrix} \psi'_+ \\ \psi'_- \end{pmatrix} = \begin{pmatrix} \varepsilon_0 - \varepsilon & 0 \\ 0 & \varepsilon_0 + \varepsilon \end{pmatrix} \begin{pmatrix} \psi'_+ \\ \psi'_- \end{pmatrix} \quad (18)$$

that is, as $\vec{E} = 0$, $\psi'_+ = \varphi' + \chi'$ and $\psi'_- = \varphi' - \chi'$ are two degeneration states of $\hat{H}$ with the same eigenvalue $\varepsilon_0$; as $\vec{E} \neq 0$, the degeneracy of the two-fold multiplet is broken by the electrostatic field $\vec{E}$, note that here involves the interaction $i\frac{e}{2m}\vec{\sigma}\cdot\vec{E}$ (associated with *spin-like* degrees of freedom) rather than the interaction $\frac{e}{2m}\vec{\sigma}\cdot\vec{B}$ (associated with spin degrees of freedom). Therefore, Equ. (18) shows that some specific *electrostatic* fields can also affect the spin quantum states of a *resting* electron by means of the interaction $i\frac{e}{2m}\vec{\sigma}\cdot\vec{E}$ that related to *spin-like* degrees of freedom.

## 5. Potential applications of interactions related to the generators of Lorentz boost

Conventionally electronics only sensitive to electron's charge, spin degree of freedom ignored. Spintronics, a new research field develop in recent years, is based on the up (↑) or down (↓) spin of carriers rather than on electrons or holes as in traditional semiconductor electronics, and involves the study of active control and manipulation of spin degrees of freedom in solid-state systems [9]. With the passage of time, the size of electronic devices becomes smaller and smaller, and the quantum-mechanics effects become more and more significant. Further, as the size of some future electronic devices becomes so small that the effects of relativistic quantum mechanics and quantum field theory, play an important role in determining the properties of these electronic devices, and are crucial for their applications, we would have to investigate these device technologies based on relativistic quantum mechanics and quantum field theory rather than nonrelativistic quantum mechanics.

In fact, though most microscopic interactions in condensed matter physics can be accurately described by nonrelativistic quantum mechanics, spin-orbit coupling that arises from the relativistic-quantum-mechanics effects described by the Dirac equation, is crucial for spintronics device applications [10], that is, one can make use of the spin-orbit coupling to manipulate electron spins by purely electric means [10-21]. For example, the Refs. [13-14] show that a purely electrostatic field can affects the spin quantum states of a moving electron, i.e. spin and motional degrees of freedom are coupled via spin-orbit coupling, which due to the fact that an electric field $\vec{E}$ in lab frame causes a magnetic field $\vec{B} = \vec{E} \times \vec{v}$ in rest frame of moving electron with the velocity $\vec{v}$.

However, as mentioned in section 4, some purely *electrostatic* fields can affect the spin quantum states of a *resting* electron by means of the interaction related to *spin-like* degrees of freedom, this may open a new pathway for spintronics to the manipulation of electron spins in the absence of applied magnetic fields. That is, as for future spintronics devices with smaller size, not only spin degrees of freedom related to the spin matrices $\bar{\Sigma}$, but also *spin-like* degrees of freedom related to the spin matrices $\bar{K}$ may enter into our consideration. In other words, a complete study for spintronics should involve the investigation of active control and manipulation of 4D spin-tensor $S^{\mu\nu}$ degrees of freedom via the electromagnetic field tensor $F_{\mu\nu}$, where involves the coupling between the 4D spin and the orbit-angular-momentum tensors. Then the interactions related to the generators of Lorentz boost may be valuable in future technologies.

As an example, we come back to the last two terms on the right-hand side of Equ. (13). When the electric and magnetic field strengths $\vec{E}$ and $\vec{B}$ as two independent external fields satisfy $\vec{E} = k\vec{B}$ ( $k$ is a constant), we have $[\vec{\Sigma} \cdot \vec{B}, \vec{K} \cdot \vec{E}] = 0$, that is, $g_s \frac{e}{2m} \vec{\Sigma}$ and $g_s \frac{e}{2m} \vec{K}$ have common eigenstates. This means that we can simultaneously make use of the electric and magnetic field strengths $\vec{E}$ and $\vec{B}$ to orientate an electron (e.g. the directions of $\vec{E}$ and $\vec{B}$ are both perpendicular to a two-dimensional plane where the polarization electrons are localized and orientated). This offers other ways to strengthen the control and manipulation for the spin degrees of freedom of electron and may open attractive and multiplicate applications in future potential electronic devices with smaller size.

## 6. Conclusions

Up to now, we have obtained the following conclusions:

1). Starting from the Dirac equation of electron in external fields, one can obtain not only the usual 3D spatial spin-orbit coupling, but also the coupling between the 4D spin and orbit-angular-momentum tensors, and the latter includes both the usual 3D spatial spin-orbit coupling and the interactions related to the Lorentz boost vectors.

2). To a charged system with nonzero Lorentz boost vectors, there may correspond to an electric moment, just as that, to a charged system with nonzero 3D spatial angular momentum, there may correspond to a magnetic moment. Especially, corresponding to the infinitesimal generators of Lorentz boost (i.e. the *spin-like* matrices) of electron, an electron has the *spin-like* degrees of freedom that belong to *the particle-antiparticle space* spanned by the positive- and negative-energy states of electron (it happened that a superposition of plane waves of positive- as well as of negative-energy is necessary to obtain a wave packet of electron), and this offer an *induced* electric moment for the electron.

3). By the *spin-like* degrees of freedom of electron, some specific *electrostatic* fields can even affect the spin quantum states of a *resting* electron, which involve the interaction between an *electrostatic* field and the *induced* electric moment of electron. Furthermore, when external magnetic and electric fields are parallel (hence independent), one can simultaneously make use of them to act on the spin and the *spin-like* degrees of freedom of electron, by which to fix on the spin orientation of electron.

4). The investigations for the interactions related to the generators of Lorentz boost, may offer some new pathways for active control and manipulation of the spin degrees of freedom of electron in future quantum information and quantum computation devices with smaller size.

**Acknowledgments**

This work was supported by China National Natural Science Foundation 69971008 and the Excellent Young Teachers Program of MOE of China.

## Appendix A

## Physical meaning of the Lorentz boost vectors

In general, a 3D orbital angular momentum $\hat{L} = \bar{x} \times \hat{p}$ can be regarded as a 3D *momentum moment*, as we extend this to the 4D case, we have $\bar{x} \to x^\mu$, $\hat{p} \to \hat{p}^\mu$, $\hat{L} \to \hat{L}^{\mu\nu} = x^\mu \hat{p}^\nu - x^\nu \hat{p}^\mu$, and $\hat{L}^{\mu\nu}$ can be regarded as a 4D *momentum moment*.

In an analogous manner (see for example, Ref. [22]), we can extend the traditional relation between 3D angular momentum and magnetic moment to the 4D case. As for the electromagnetic 4-potential $A_\mu(x)$, we have

$$A^\mu(\vec{x},t) = \frac{d}{dt'}\int A^\mu(\vec{x},t)dt' = \frac{1}{4\pi}\int \frac{J^\mu(\vec{x}',t')}{|\vec{x}-\vec{x}'|}d^3x'$$

$$= \frac{1}{|\vec{x}|}\int J^\mu(\vec{x}',t')d^3x' + \frac{\vec{x}}{|\vec{x}|^3}\int J^\mu(\vec{x}',t')\vec{x}'d^3x' + \cdots\cdots ,$$  (A-1)

where $t' = t - r, r = |\vec{x} - \vec{x}'|$ ($\hbar = c = 1$), $J^\mu$ is a localized divergenceless current, which permits simplification and transformation of the expansion (A-1). Let $f(x')$ and $g(x')$ be well-behaved functions of $x'$ to be chosen below

$$\int(fJ\cdot\Delta g + gJ\cdot\Delta f)d^4x = 0 \qquad (\Delta\cdot J = 0),$$  (A-2)

where $\Delta$ denotes the 4D gradient operator. Let $f = x_\mu$ and $g = x_\nu$, we have

$$\int(x_\mu J_\nu + x_\nu J_\mu)d^4x = 0,$$  (A-3)

$$\vec{x}\cdot\frac{d}{dt'}\int \vec{x}'J_\mu(x')d^4x' = \sum_j x_j \frac{d}{dt'}\int x'_j J_\mu(x')d^4x'$$

$$= -\frac{1}{2}\sum_j x_j \frac{d}{dt'}\int (x'_i J_\mu - x'_\mu J_j)d^4x'$$  (A-4)

It is customary to define *the 4D electromagnetic moment density tensor*

$$m^{\mu\nu} = \frac{1}{2}[x^\mu J^\nu - x^\nu J^\mu],$$  (A-5)

and its integral as the electromagnetic moment (not a tensor)

$$M^{\mu\nu} = \frac{1}{2}\int[x'^\mu J^\nu - x'^\nu J^\mu]d^3x'.$$  (A-6)

Assuming that $J^\mu$ is provided by $N$ charged particles with momenta $p_n^\mu = mu_n^\mu$ ($n = 1,2,...,N$, $m$ is the proper mass, $u_n^\mu$ is the 4-velocity of the $n$-th particle) and charges $e$, then $J^\mu(x') = \sum_n eu_n^\mu(t')\delta^3(\vec{x}-\vec{x}'_n)\frac{d\tau}{d\tau'}$, where $\tau$, $\tau'$ are the proper times. When $t = t'$, we have

$$M^{\mu\nu} = \frac{e}{2m'}L^{\mu\nu} = \frac{e}{2m}\sum_n(x_n^\mu p_n^\nu - x_n^\nu p_n^\mu)\frac{d\tau}{dt}$$  (A-7)

where $m'$ is the relativistic mass, $L^{\mu\nu} = \sum_n(x_n^\mu p_n^\nu - x_n^\nu p_n^\mu)$ is the 4D total orbital-angular-momentum tensor.

In a word, when we consider the multipole expansion of the electromagnetic 4-potential of a charged system, a 4D angular momentum tensor implies an electromagnetic moment that is assembled into by a magnetic moment and an electric moment components, where the 3D angular momentum means a magnetic moment, while the Lorentz boost vector implies the electric moment. For these a heuristic understanding or an intuitive physical picture can be obtained from Ref. [23].

## Appendix B

## Small Components of the Wave Function of Electron

In appendix B, *via* an analogy between the free electron and free electromagnetic field, we study the relation between the large components and the small components of the wave function of electron, and show when the small components can not be ignored. A free electron with mass $m$ is described by the Dirac equation for spin-$1/2$ particles:

$$(i\gamma^\mu \partial_\mu - m)\psi(x) = 0 \text{, or } i\partial_t \psi(x) = \hat{H}\psi(x) \tag{B-1}$$

where $\hat{H} = \vec{\alpha}\cdot\hat{p} + \gamma^0 m$ is the Hamiltonian, $\gamma^\mu$ are the four $4\times 4$ Dirac matrices and $\vec{\alpha} = \beta\vec{\gamma}$, with the help of the Pauli matrix vector $\vec{\sigma} = (\sigma_1, \sigma_2, \sigma_3)$

$$\sigma_1 = \begin{pmatrix} 0 & 1 \\ 1 & 0 \end{pmatrix}, \ \sigma_2 = \begin{pmatrix} 0 & -i \\ i & 0 \end{pmatrix}, \ \sigma_3 = \begin{pmatrix} 1 & 0 \\ 0 & -1 \end{pmatrix} \tag{B-2}$$

we have

$$\gamma^0 = \begin{pmatrix} I_{2\times 2} & 0 \\ 0 & -I_{2\times 2} \end{pmatrix}, \ \vec{\gamma} = \begin{pmatrix} 0 & \vec{\sigma} \\ -\vec{\sigma} & 0 \end{pmatrix}, \ \vec{\alpha} = \begin{pmatrix} 0 & \vec{\sigma} \\ \vec{\sigma} & 0 \end{pmatrix} \tag{B-3}$$

where $I_{n\times n}$ is the $n\times n$ unit matrix, $n = 2, 3, 4...$.

On the other hand, in the vacuum and source-free medium, the electric field intensity $\vec{E}$ and the magnetic field intensity $\vec{B}$ satisfy the Maxwell equations

$$\nabla \times \vec{E} = -\partial_t \vec{B}, \nabla \times \vec{B} = \partial_t \vec{E} \qquad (B\text{-}4)$$

while the transversality conditions $\nabla \cdot \vec{E} = 0, \nabla \cdot \vec{B} = 0$ are contained in Equ. (B-4). The Maxwell equations (B-4) can be rewritten as the *Dirac-like form*

$$i\beta^\mu \partial_\mu \Theta(x) = 0, \text{ or } i\partial_t \Theta(x) = \hat{H}_{em} \Theta(x) \qquad (B\text{-}5)$$

where $\hat{H}_{em} = \vec{\theta} \cdot \hat{p}$ is the Hamiltonian, $\beta^\mu$ are the four $6 \times 6$ matrices and $\vec{\theta} = \beta^0 \vec{\beta}$. With the help of the matrix vector $\vec{\tau} = (\tau_1, \tau_2, \tau_3)$

$$\tau_1 = \begin{pmatrix} 0 & 0 & 0 \\ 0 & 0 & -i \\ 0 & i & 0 \end{pmatrix}, \tau_2 = \begin{pmatrix} 0 & 0 & i \\ 0 & 0 & 0 \\ -i & 0 & 0 \end{pmatrix}, \tau_3 = \begin{pmatrix} 0 & -i & 0 \\ i & 0 & 0 \\ 0 & 0 & 0 \end{pmatrix} \qquad (B\text{-}6)$$

we have

$$\beta^0 = \begin{pmatrix} I_{3\times 3} & 0 \\ 0 & -I_{3\times 3} \end{pmatrix}, \vec{\beta} = \begin{pmatrix} 0 & \vec{\tau} \\ -\vec{\tau} & 0 \end{pmatrix}, \vec{\theta} = \begin{pmatrix} 0 & \vec{\tau} \\ \vec{\tau} & 0 \end{pmatrix} \qquad (B\text{-}7)$$

$$\eta = \begin{pmatrix} E_1 \\ E_2 \\ E_3 \end{pmatrix}, \rho = \begin{pmatrix} iB_1 \\ iB_2 \\ iB_3 \end{pmatrix}, \Theta = \begin{pmatrix} \eta \\ \rho \end{pmatrix} \equiv \begin{pmatrix} \vec{E} \\ i\vec{B} \end{pmatrix} \qquad (B\text{-}8)$$

By applying Equ. (B-5)-(B-8), we have

$$(\vec{\tau} \cdot \nabla)\rho = -\partial_t \eta, \quad (\vec{\tau} \cdot \nabla)\eta = -\partial_t \rho \qquad (B\text{-}9)$$

Equ. (B-9) is the matrix form of Equ. (B-4).

Similarly, in Equ. (B-1), let the four-component spinor $\psi$ is decomposed into two two-component spinors $\varphi$ and $\chi$, i.e., $\psi = \begin{pmatrix} \varphi \\ \chi \end{pmatrix}$, in terms of the Pauli matrices, the Dirac equation (B-1) can also be rewritten as the *Maxwell-like form*

$$(\vec{\sigma} \cdot \nabla)\chi = (-\partial_t + im)\varphi, \quad (\vec{\sigma} \cdot \nabla)\varphi = (-\partial_t - im)\chi \qquad (B\text{-}10)$$

By the way, Let $\lambda_e = 1/2$ and $\lambda_p = 1$, $\hat{L} = \bar{x} \times \hat{p}$ be the orbital angular momentum of particles, one has $[\hat{H}, \hat{L} + \lambda_e \vec{\Gamma}] = 0$ and $[\hat{H}_{em}, \hat{L} + \lambda_p \vec{\Lambda}] = 0$, where $\vec{\Gamma} = \begin{pmatrix} \bar{\sigma} & 0 \\ 0 & \bar{\sigma} \end{pmatrix}$, $\vec{\Lambda} = \begin{pmatrix} \bar{\tau} & 0 \\ 0 & \bar{\tau} \end{pmatrix}$, $\lambda_e \vec{\Gamma} \cdot \lambda_e \vec{\Gamma} = \frac{1}{2}(\frac{1}{2}+1) I_{4 \times 4}$ and $\lambda_p \vec{\Lambda} \cdot \lambda_p \vec{\Lambda} = 1(1+1) I_{6 \times 6}$, then $\lambda_e = 1/2$, $\lambda_p = 1$ denote the spins of electron and photon, respectively, and $\lambda_e \vec{\Gamma}$, $\lambda_p \vec{\Lambda}$ denote the corresponding spin-matrix operators. In fact, let $\bar{p}$ be the momentum of particles, we have $\frac{1}{|\bar{p}|} \bar{p} \cdot \lambda_e \vec{\Gamma} \psi = \pm \lambda_e \psi$, $\frac{1}{|\bar{p}|} \bar{p} \cdot \lambda_p \vec{\Lambda} \Theta = \pm \lambda_p \Theta$, that is, $\lambda_e \vec{\Gamma} \cdot \frac{\bar{p}}{|\bar{p}|}$ and $\lambda_p \vec{\Lambda} \cdot \frac{\bar{p}}{|\bar{p}|}$ are the helicity operators of electron and photon, respectively. Moreover, $\vec{\alpha} = \begin{pmatrix} 0 & \bar{\sigma} \\ \bar{\sigma} & 0 \end{pmatrix}$ and $\vec{\theta} = \begin{pmatrix} 0 & \bar{\tau} \\ \bar{\tau} & 0 \end{pmatrix}$ are the velocity operators of electron [5] and photon, respectively, e.g., $\Theta^+ \vec{\theta} \Theta = 2 \bar{E} \times \bar{B}$, where $\Theta^+$ is the Hermitian adjoint of $\Theta$ (and so on).

That is to say, the Maxwell equations for free electromagnetic fields $\Theta = \begin{pmatrix} \bar{E} \\ i\bar{B} \end{pmatrix}$ can be rewritten as the *Dirac-like equation*, conversely the Dirac equation for free electrons $\psi = \begin{pmatrix} \varphi \\ \chi \end{pmatrix}$ can be rewritten as the *Maxwell-like equations*. Comparing the Dirac equation (B-1) with the *like-Dirac* equation (B-5), or the *Maxwell-like* equation (B-10) with the Maxwell equation (B-9), we can show that, $\varphi$ is to $\chi$, as $\bar{E}$ is to $\bar{B}$. Furthermore, we have the following analogies:

(1) By Equ. (B-9) and $\Theta = \begin{pmatrix} \eta \\ \rho \end{pmatrix} \equiv \begin{pmatrix} \bar{E} \\ i\bar{B} \end{pmatrix}$, a moving or time-varying electric field $\bar{E}$ induce the magnetic field $\bar{B}$ and vice versa; By Equ. (B-10) and $\psi = \begin{pmatrix} \varphi \\ \chi \end{pmatrix}$, a moving or time-varying component $\varphi$ of $\psi$ induce the component $\chi$ and vice versa.

(2) Under the exchange $\vec{E} \leftrightarrow i\vec{B}$, the free electromagnetic fields have the electricity-magnetism duality property; Under the exchange $\varphi \leftrightarrow \chi$, the free Dirac fields have particle-antiparticle symmetry property.

(3) The quantity $\vec{E}^2 - \vec{B}^2$ ($= \bar{\Theta}\Theta \equiv \Theta^+ \beta^0 \Theta$) is Lorentz invariant; The quantity $\varphi^+\varphi - \chi^+\chi$ ($= \bar{\psi}\psi \equiv \psi^+\gamma^0\psi$) is also Lorentz invariant.

(4) The quantity $\vec{E}^2 + \vec{B}^2$ ($= \Theta^+\Theta$) is proportional to the density of photon number; The quantity $\varphi^+\varphi + \chi^+\chi$ ($= \psi^+\psi$) is proportional to the density of electron number.

(5) For the field $\Theta = \begin{pmatrix} \vec{E} \\ i\vec{B} \end{pmatrix}$ of *electric* charge, the electric field $\vec{E}$ is the large component and the magnetic field $\vec{B}$ the small component, while for the field $\Theta = \begin{pmatrix} \vec{E} \\ i\vec{B} \end{pmatrix}$ of *magnetic* charge, the magnetic field $\vec{B}$ is the large component and the electric field $\vec{E}$ the small component. Analogously, for the field $\psi = \begin{pmatrix} \varphi \\ \chi \end{pmatrix}$ of *negative* electron, $\varphi$ is the large component and $\chi$ the small component, while for the field $\psi = \begin{pmatrix} \varphi \\ \chi \end{pmatrix}$ of *positive* electron, $\chi$ is the large component and $\varphi$ the small component.

In view of the statement (1) above, i.e., a moving or time-varying component $\varphi$ induce the component $\chi$ and vice versa, as $|\vec{p}| \sim m$, the small component can not to be ignored; likewise, as the size of the wave packet $r \sim 1/m$ (e.g., owing to a strong external field), because of the Heisenberg uncertainty principle, the small component would have an appreciable effect. In a word, when a wave packet of electron is moving with high speeds or varies rapidly, or its size is sufficiently small, or in the present of a strong electromagnetic field, its small components and the related effects can not be ignored. Especially, consider that the negative-energy solutions of the Dirac equation correspond to the positive-electron ones, and in these wavepackets the relative intensities of positive- and negative-energy components are proportional to $|\varphi|^2/|\chi|^2$, then a negative-electron wavepacket does contain a positive-electron component (the latter as the small component of the wavepacket, is the relativistic effect of the large component of the wavepacket), and vice versa, just as that a moving or time-varying electric field is always accompanied by a magnetic field component, and vice versa.

From the point of view of mathematics, the fact that free electromagnetic field can be also described as a $6\times1$ spinor $\Theta = \begin{pmatrix} \vec{E} \\ i\vec{B} \end{pmatrix}$, lies in that the free electromagnetic field corresponds to both the $(\frac{1}{2},\frac{1}{2})$ representation of the Lorentz group via the electromagnetic 4-vector $A_\mu$ and the $(1,0)\oplus(0,1)$ representation via the electromagnetic field $F_{\mu\nu} \leftrightarrow \vec{E},\vec{B}$.